\documentclass[amsmath,amssymb,aps,pre,twocolumn,superscriptaddress,floatfix]{revtex4-1}

\usepackage{graphicx}
\usepackage{dcolumn}
\usepackage{bm}
\usepackage{hyperref}
\usepackage{changes} 
\usepackage[utf8]{inputenc}
\usepackage{fancyhdr}
\definechangesauthor[color=purple, name={Paul Richter}]{P}
\usepackage{comment}
\usepackage{listings}
\usepackage{lineno}
\usepackage{chemfig}
\usepackage{dsfont}
\usepackage{soul,color}
\usepackage{environ}
\usepackage{bbold}
\usepackage{csquotes}
\usepackage{comment}
\begin{document}
\renewcommand{\vec}[1]{\mathbf{#1}}
 
\newcommand*{\addheight}[2][.5ex]{\raisebox{0pt}[\dimexpr\height+(#1)\relax]{#2}}
 
\newcommand\size{0.9}
 
\setlength{\marginparwidth}{2cm}

\title{Large-scale GPU-based network analysis of the human T-cell receptor repertoire}

\author{Paul Richter}
\affiliation{Institute for Theoretical Physics, ETH Zurich, 8093, Zurich, Switzerland}

\newcommand{\CDR}{CDR3$\beta$~} 

\date{\today}
\begin{abstract}
Understanding the structure of the human T-cell receptor  repertoire is a crucial precondition to understand the  ability of the immune system to recognize and respond to antigens. T cells are often compared via the complementarity determining region 3 (CDR3) of their respective T-cell receptor $\beta$ chains. 
Nevertheless, previous studies often simply compared if \CDR sequences were equal, while 
network theory studies were usually  limited to several ten thousand sequences due to the high  computational effort of constructing the network. 
To overcome that hurdle, we introduce the GPU-based algorithm {\sc TCR-NET} to construct large-scale \CDR similarity 
networks using model-generated and empirical sequence data with up to 800,000 
\CDR sequences on a normal computer for the first time. Using network analysis methods we study the structural properties of these  
networks and conclude that (i) the fraction of public TCRs depends on the size of the TCR repertoire, along with the exact (not  unified) definition of \enquote{public} sequences, (ii) the TCR network
 is assortative with the average neighbor degree being proportional to the squareroot of the degree of a node and 
(iii) the repertoire is robust against 
 losses of TCRs. 
 Moreover, we analyze the networks of antigen-specific TCRs for different antigen families  and find differing clustering coefficients and assortativities. 
TCR-NET offers better access to assess large-scale TCR repertoire networks,   opening the possibility to quantify 
their structure and quantitatively distinguish their ability to react to antigens, which we anticipate to become a useful tool in a time of increasingly large amounts of  
 repertoire sequencing data becoming available.
\end{abstract}

\newcommand{\NSeqPublicAnalysis}{$N = 8\times 10^{5}$ } 
\newcommand{\NSeqRobustness}{$N = 8\times 10^{5}$ } 
\newcommand{\NSeqPhaseTransition}{$N = 8 \times 10^{3}$ } 
\newcommand{\ie}{\textit{i.e.}} 
\newcommand{\eg}{\textit{e.g.}}

\newcommand{\NSeqCliques}{$4.8\times 10^{4}$ }
\newcommand{\sigmacmd}{$a$}

\newcommand{\masterscale}{0.8} 
\newcommand{\itcmd}{\itshape}

\newcommand{\rmcmd}{\mathrm}
\newcommand{\mathbbcmd}{\mathbbm}

\newcommand{\ro}[4][]{
  \tikz\node [circle, minimum width = #2,
    path picture = {
      \node [#1] at (path picture bounding box.center) {
        \includegraphics[width=#3]{#4}};
    }] {};}

\maketitle

\maketitle

In order to protect the body against a wide range of pathogens, 
the immune system maintains a high diversity of 
T-cell receptors (TCRs)~(\cite{e2003fundamental} p. 397). This happens via somatic recombination of TCR genes in the thymus, known as V(D)J recombination. While theoretically, more than $10^{15}$ different TCRs ~\cite{miles2011bias, laydon2015estimating} could be formed, the real diversity of TCRs is estimated to be  about $10^8$ in humans~\cite{soto2020high} and about $10^6$ in mice~\cite{casrouge2000size}. 

Within the TCR gene, the antigen specificity is mainly determined by the complementarity determining region 3 (CDR3) in the variable domain (not to be confused with the variable region). Moreover, in 95 \% of human T cells the TCR consists of an $\alpha$ chain and a $\beta$ chain~(\cite{e2003fundamental} p. 314). 
While up to one third of all T cells express two different TCR$\alpha$ chains~\cite{padovan1993expression, schuldt2019dual}, only one percent express two different 
TCR$\beta$ chains~\cite{padovan1995normal, schuldt2019dual}. Hence, the diversity of TCR$\beta$ chains is closely related to the diversity of T cells. Consequently, we
will examine CDR3 regions from TCR$\beta$ chains (\CDR sequences)  throughout this analysis. 

Studying the similarity landscapes of \CDR sequences~\cite{miho2019large} therefore is of general interest in order to understand  the architecture of the TCR repertoire. 
Network analysis offers a way to compare \CDR sequence pools and has previously been employed in biology~\cite{pavlopoulos2011using, GOSAK2018118, emmert2011networks}, notably by helping to identify  functional relations in metabolic and protein-protein interaction (PPI) 
networks~\cite{LI2011143, 6081844,  pieroni2008protein, raman2010construction}.
Moreover, network science methods found applications in studying the evolution of different genes~\cite{novak2010graph}, drug discovery \cite{doi:10.1038/clpt.2013.176}, brain research \cite{doi:10.1089/brain.2011.0055}, and in discovering biomarkers~\cite{doi:10.1517/17530059.2012.718329}.

Advances in repertoire sequencing technology and the resulting increased availability of 
large-scale sequencing data led to  the application of network theory to  immune repertoires~\cite{ben2011whole, chang2016network, hoehn2015dynamics, lindner2015diversification, madi2017t, bashford2013network,miho2019large}. 
Nevertheless, with exception of Ref. \cite{miho2019large}, that employed a large-scale computer cluster in order to analyze an antibody repertoire containing more than $10^5$ \CDR sequences, networks were constructed based on no more than from a few hundred to several ten thousand sequences. 
Each node (also referred to as a {\it vertex}) in an immune repertoire network represents a clone and pairs of nodes are connected if the corresponding amino acid (a.a.) sequences are sufficiently similar (we will compare them by calculating their Levenshtein distance). 
Here, we use an immune-repertoire network approach to analyze the structural properties of large-scale TCR \CDR~repertoires as networks  (hence TCR networks for simplicity). 

\begin{figure*}
    \centering
    \includegraphics[width=0.9\textwidth]{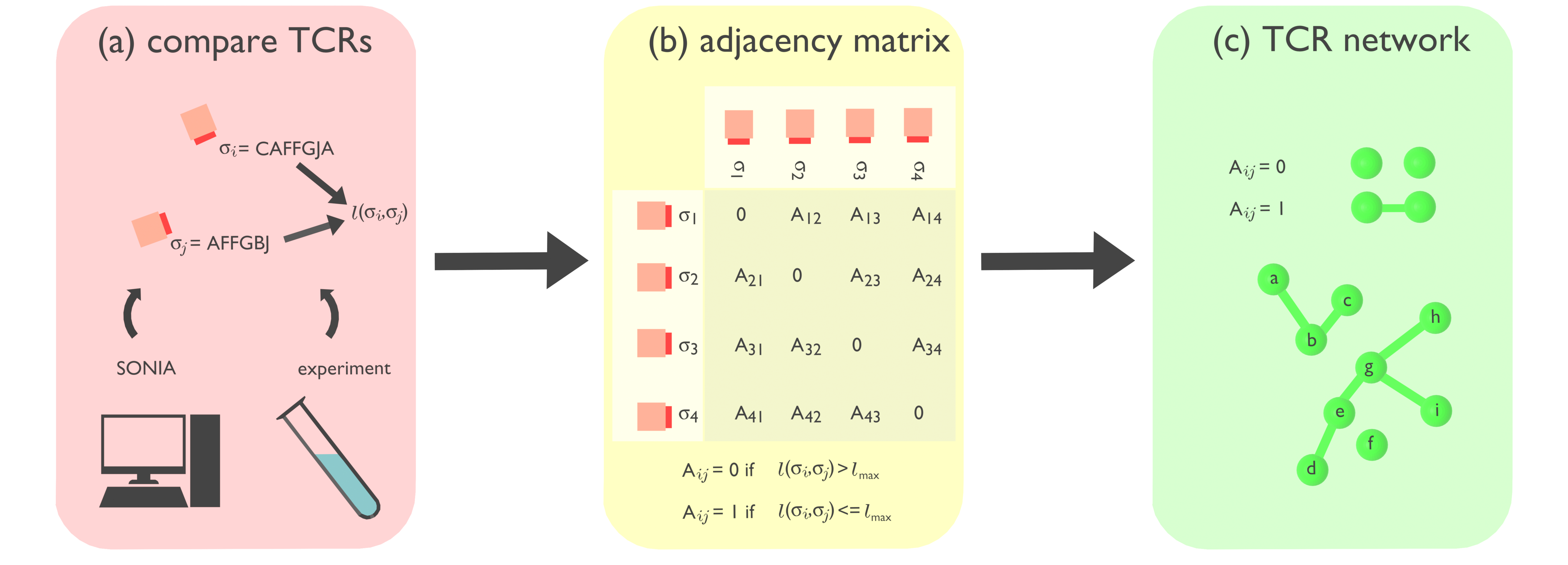}
    \caption{{\bf Scheme of the TCR analysis framework. }(a) 
    Empirical or generated TCRs are compared by  determining the string distance 
    (throughout our analysis we choose the Levenshtein distance $l(\sigma_i, \sigma_j)$, as done previously \cite{miho2019large})
    between their respective \CDR amino acid sequences $\sigma_i$ and $\sigma_j$. If their string distance does not exceed a threshold value $l_{max}$, we define them to be {\it adjacent}, otherwise {\it non-adjacent}. 
    (b) Among large numbers of \CDR sequences the adjacencies 
    between all possible \CDR sequence pairs can be described with the adjacency matrix. 
    (c) Nodes in a TCR network represent specific 
    \CDR 
    sequences, while edges between nodes indicate that 
    they are adjacent, i.e., their string distance does
    not exceed $l_{max}$. 
    }
    \label{fig:algorithm}
\end{figure*}

In order to construct larger TCR networks in a reasonable amount of time, we introduce the GPU-based network construction framework  TCR-NET, which can create a TCR network from 800,000 \CDR sequences in less than an hour on a normal computer (for details about the GPU implementation and performance see Methods section \ref{sec:GPU_algorithm}). Fig.~\ref{fig:algorithm} schematically displays  the steps of our TCR network construction framework. 
First, empirical or generated \CDR  sequences are loaded as an input to our framework (Fig. \ref{fig:algorithm} (a)). We employ SONIA (see Methods section \ref{sonia}) to computer-generate \CDR sequences and use a databank from Ref.~\cite{soto2020high} to 
obtain empirical \CDR sequences from naive CD8 T cells. 
Second, Levenshtein distances~\cite{levenshtein1966binary} between all pairs of \CDR sequences are 
determined, which forms the computationally most expensive task for repertoires with several thousand or million sequences. If the Levenshtein  distance between a pair of sequences exceeds a threshold value $l_{max} \in \mathbb{N}$, the 
sequences are defined as {\it non-adjacent}, otherwise they are {\it adjacent}. 
The {\it adjacency matrix} (Fig. ~\ref{fig:algorithm} (b)) stores the value 0 for all non-adjacent pairs of sequences and 1 for all adjacent pairs. 
Third, TCR networks are constructed by connecting nodes (\ie, \CDR  sequences) if their \CDR Levenshtein distances are smaller than or equal to $l_{max}$, i.e., if the sequences  are adjacent (Fig.~\ref{fig:algorithm} (c)).
A similar way to construct networks from  antibody \CDR sequences was employed in Ref. ~\cite{miho2019large}.  
We can then study the constructecd TCR networks, employing the established toolbox of network theory. Further details on the generation 
of \CDR sequences, GPU-based Levenshtein distance calculation and network analysis concepts are summarized in the Methods section. 

Since the immune repertoire aims to respond to antigens, for each antigen  there exist specific TCRs, which can detect them. 
They are called antigen cognate TCRs.
We will investigate, how the structure of antigen-cognate parts of the TCR repertoire differs among different virus families, by investigating network characteristics, such as assortativity and degree distribution for different antigens. For that we use antigen cognate \CDR sequences from a public databank~\cite{tickotsky2017mcpas}. 
\begin{figure*}[htb]
    \centering
    \includegraphics[scale = 0.8]{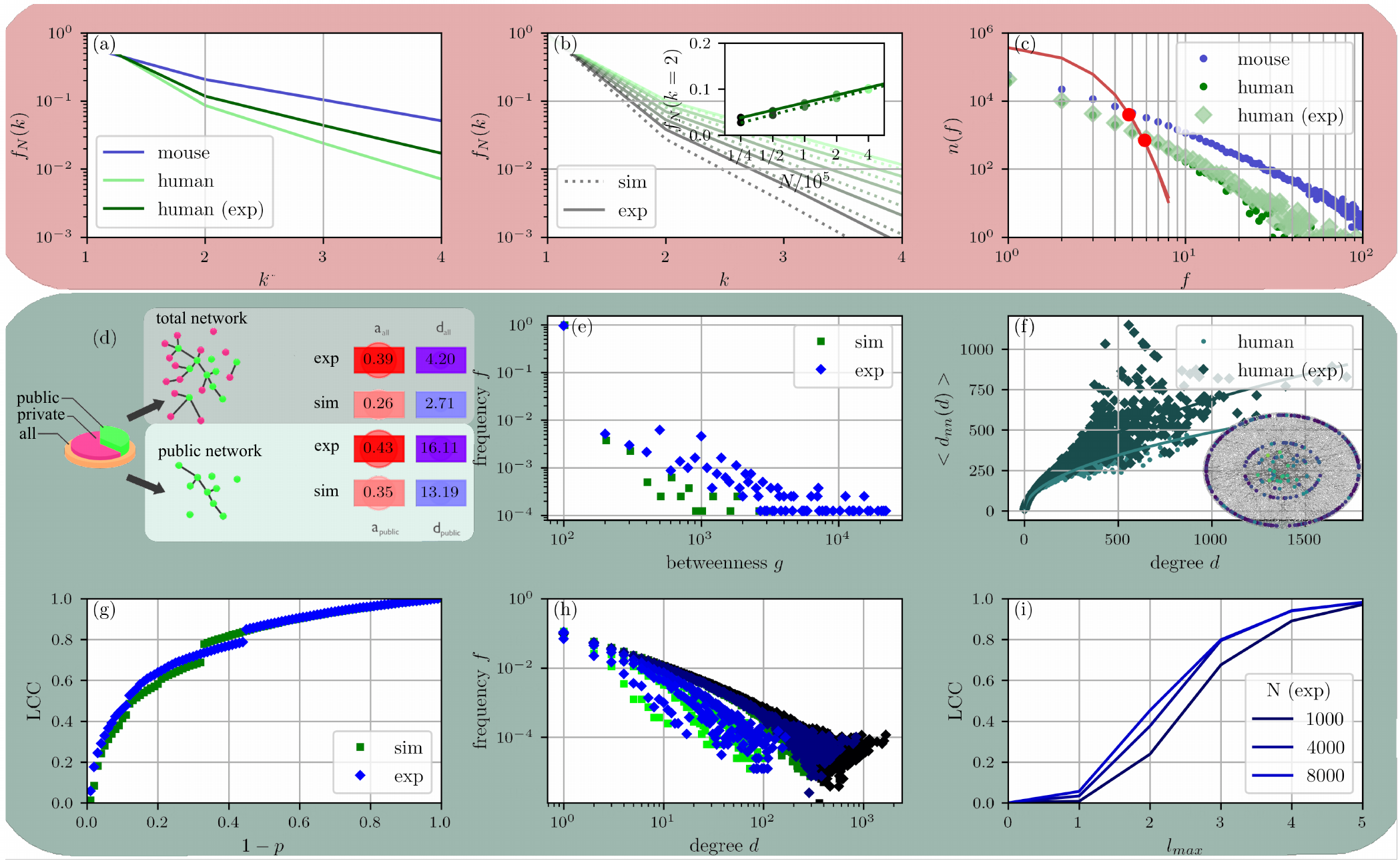}
    \caption{{\bf Analysis of \CDR sequence pools.} 
Panel (a) shows the normalized frequencies 
    $f_N(k)$ of \CDR sequences that occur in at least $k$ 
    individuals, where each individual carries $N=10^6$ \CDR  sequences.
Panel (b) shows the distribution $f_N(k)$ for different repertoire sizes $N \in 
    \{1/4\times 10^{5},
    1/2\times 10^{5}, 1\times 10^{5}, 2\times 10^{5}, 4\times
    10^{5}\}$. In the inset of panel (b) we plot $f_N(k=2)$ in
    dependence of the repertoire size $N$  in a logarithmic plot and compare it with a  logarithmic curve.  
In panel (c) we plot the number of different \CDR sequences with a certain      
    frequency $n(f)$ and compare the curve from a TCR repertoire of size $N=10^{6}$ (blue and green) with the red curve of a random selection (for details please refer to the text). 
Panel (d) shows the assortativity and the mean degree for the total TCR network 
    and the public TCR network respectively, constructed based on four \CDR sequence pools of $8\times 10^3$ \CDR sequences each and the threshold $l_{max} = 2$. 
In panel (e) we plot the distribution of betweenness centralities for $N = 8\times 10^3$ and  $l_{max} = 1$. 
In panel (f) the average neighbor
    degree of a node is shown in dependence of the degree of the
    node itself for $N=8\times 10^5$ and $l_{max}=1$. The monotonically increasing function, indicates an assortative TCR network.  
    {\bfseries Robustness analysis.} 
Panel (g) shows the relative size of the largest connected component (LCC) as a 
    function of $1-p$, where $p$ is the fraction of removed nodes  (\NSeqRobustness,  $l_{max} = 1$). 
    The TCR network does not exhibit a  
    percolation phase transition and is robust against node removal. (h) The corresponding degree distribution of the same network is plotted for $p\in \{0.01, 0.49, 0.89, 0.95, 0.98\}$ with lighter 
    green (blue for empirical data) indicating a larger $p$. With increasing $p$, the decline  
     becomes steeper as the number of nodes decreases. 
    {\bfseries Connectivity transition in TCR networks.} 
    In panel (i) we show the relative size of the LCC in dependence of $l_{max}$ for different TCR repertoire sizes $N$.
    A transition can be observed, which shifts towards lower $l_{max}$ for higher repertoire sizes.
}
\label{fig:fig_2}
\end{figure*}

\section*{Results}

In section \ref{public}, we will analyze  the diversity of \CDR sequences as well as the fraction of public sequences in the TCR repertoire.  
In section \ref{network_structure}, we will investigate network properties of the TCR repertoire, namely the average neighbor degree distribution and  the robustness of the system against 
removal of T cells. In section \ref{virus}, we will finally investigate  the network properties of antigen-cognate \CDR sequences and determine the a.a. composition of the \CDR sequences.
\subsection{Public sequences}
\label{public}
During V(D)J recombination each unique  \CDR sequence has a certain probability of being generated. 

Thus, each individual has a probability $p_{ind}(\boldsymbol{\sigma})$ of possessing a TCR with a unique \CDR sequence  $\boldsymbol{\sigma}$, which can be high or low. 
Hence, we do not just want to know whether a sequence occurs in 
only one individual (private) or in every individual (public). 
Instead, the possibility arises that the sequence exists in multiple, but not 
all individuals. That prompts the question whether these sequences, 
which occur in multiple but not all individuals, 
are private or public. This topic was investigated in
previous publications~\cite{elhanati2018predicting, quigley2010convergent, robins2009comprehensive, freeman2009profiling, benichou2012rep, six2013past, robins2010overlap, venturi2011mechanism, elhanati2014quantifying, zvyagin2014distinctive,  pogorelyy2017persisting, madi2014t, soto2020high}, still  
 there are different definitions of public, ranging from 
 defining any sequence that occurs twice or more~\cite{miho2019large} 
 to restricting the public category to sequences which occur in  
 all individuals~\cite{madi2017t}. 
In order to obtain a full view, Fig. \ref{fig:fig_2} (a) shows 
the frequency distribution of sequences, as explained below. 
\begin{figure}
    \includegraphics[scale = 1.4]{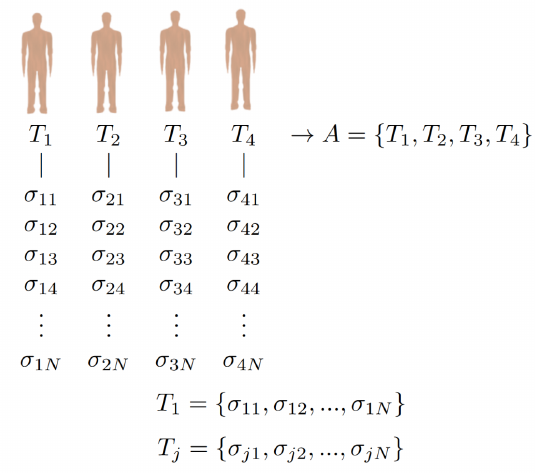}
    \caption{{\bf Conceptual illustration of the sharing number analysis for $n=4$.} Each human 
    $i$ possesses a unique TCR repertoire with a set $T_i$ of N \CDR  sequences, labeled by $\sigma_{ij}$ ($j^{th}$ sequence of the $i^{th}$ human). The set $A = \{T_1, ..., T_4\}$ of sets contains all TCR repertoires.}
    \label{fig:public_scheme}
\end{figure}

\subsubsection{Frequency decreases with sharing number} 
May the set of $n$ TCR repertoires, each from a unique individual,  be represented as $A = \{ T_i~|~0<i\leq n \}$ (see Fig. \ref{fig:public_scheme}), where $T_i = \{ {\boldsymbol \sigma_{ij} ~|~ 0<j\leq N}\}$ is a TCR repertoire comprising $N$ sequences  and we assume that all TCR repertoires $T_i$ have the same size $N$, where {\it size} of a repertoire is defined as the number of \CDR sequences within the repertoire. 

If for a given  sequence $\boldsymbol \sigma_{ij}$  another sequence $\boldsymbol \sigma_{i'j'}$ exists with $\boldsymbol \sigma_{ij} = \boldsymbol \sigma_{i'j'}$ and $i \neq i'$, that means 
that this sequence occurs in two different individuals. 
Hence, according to the 
definition in Ref.~\cite{miho2019large}, it is public. Let us define the {\it sharing number} 
$n_{{\boldsymbol \sigma}}$ as the number 
of individuals in which a sequence equal to ${\boldsymbol \sigma}$ can be found~\cite{elhanati2018predicting}.
If $n_{{\boldsymbol \sigma}} > 1$, the sequence is public, otherwise private. Moreover, 
let us define $M_k$ as the number of different sequences $\sigma_{ij} \in A$ \footnote{strictly speaking $\sigma_{ij}\in \bigcup A := \{{\boldsymbol  \sigma}|\exists T \in A: {\boldsymbol \sigma} \in T\}$}
with
$n_{{\boldsymbol \sigma_{ij}}} \geq k$, where $k$ may be any non-negative integer. 
In panel (a) of Fig. \ref{fig:fig_2} the normalized frequency $f_N(k) := \frac{M_k}{M_1}$ is plotted in dependence of $k$ on a logarithmic scale for $n=4$ 
TCR repertoires, each comprising $N=10^6$ \CDR sequences, generated by SONIA (green for human, blue for mouse) and the same number of 
empirical human \CDR sequences. 
First of all, we observe that for humans $f_N$ declines sharply with 
increasing $k$. 
A decline can be expected since it is less probable to find a sequence in many samples than in just a few. 
The empirical data exhibits a 
slower decline, though the qualitative  behavior is similar to that of generated  
sequences, and is in qualitative agreement with the results of a previous  study~\cite{elhanati2018predicting} with more human subjects but fewer \CDR sequences per repertoire. 

\subsubsection{Sharing number depends on repertoire size}

In Fig. \ref{fig:fig_2} (b) we show the distribution $f_N(k)$ from Fig. \ref{fig:fig_2} (a) again, this time for different $N \in \{1/4\times 10^{5}, 1/2\times 10^{5}, 1\times 10^{5}, 2\times 10^{5}, 4\times 10^{5}\}$. 
The distribution shifts upwards for higher $N$. The inset of Fig. \ref{fig:fig_2} (b) plots $f_{N}(k=2)$ in dependence of $N$ with logarithmic $N$-axis.
For comparison, we show a straight line.

For empirical and generated sequences we  observe a logarithmic increase of $f_N(k=2)$ with $N$ and therefore can confirm the predictions of Ref. ~\cite{elhanati2018predicting} up to  $N=400,000$ sequences. The increase is expected to slow down and saturate for higher $N$.

In agreement with previous studies~\cite{elhanati2018predicting}, it can be concluded that the fraction of public frequencies in a TCR repertoire also depends on the size $N$ of the repertoire itself. 
\subsubsection{Selection effects}
 V(D)J recombination spawns every unique \CDR sequence ${\boldsymbol \sigma}$ with a designated probability $p$. 
 Generating a set 
of sequences by V(D)J recombination and subsequent thymus selection is thus a weighted combination $C_{{\boldsymbol p}}(N_{max}, N)$ with repetitions (for more thorough definition see sections \ref{Combination} and \ref{Combination of amino acids} under Methods), 
where $N_{max}$ denotes the number of theoretically possible different sequences (sequence types), which is estimated to 
be up to more than $10^{15}$~\cite{miles2011bias, laydon2015estimating}, 
though other estimates suggest that the real number of possibilities lies only around  $10^{6}$~\cite{robins2009comprehensive}.  
$N$ is again the number of generated sequences for a TCR repertoire and ${\boldsymbol p} = \{ p_i~|~ 0<i\leq N_{max}\}$ refers to the set of probabilities $p_i$ for generating a certain sequence type ${\boldsymbol \sigma_i}$.
 This can be compared to an unweighted combination $C_{\mathbb{1}}(N_{max}, N)$, 
 where all probabilities $p_i$ are equal. 
After a combination each sequence type $\boldsymbol{\sigma}$ occurs with a frequency $f(\boldsymbol{\sigma})$. Let's denote the number of different sequence  types $\boldsymbol{\sigma}$ that have the same frequency $f(\boldsymbol{\sigma})$ with $n(f)$.
Fig.~\ref{fig:fig_2}~(c) 
plots $n(f)$ over $f$ for a TCR repertoire of size $N=10^{6}$ (blue and green dots for SONIA generated sequences of mouse and human respectively, green diamond markers for empirical human sequences). 
The generation of this 
repertoire can be viewed as a weighted combination $C_{\boldsymbol p}(N_{max}, N=10^{6})$, where $N_{max}$ 
can range from $10^{6}$ to more than $10^{15}$.
For comparison, $N=10^{6}$  
numbers were randomly chosen from the range of the first $N_{max}=10^{6}$ natural numbers, i.e. from the set $\{i~|~ 0<i\leq 10^6\}$,  
 in an unweighted combination  $C_{\mathbb{1}}(10^{6}, 10^{6})$. 
 The resulting function $n_{\mathbb{1}}(f)$ (red curve) 
 declines significantly faster than that of the weighted 
 combination $C_{\boldsymbol p}(N_{max}, N)$. Hence, 
 $n_{\mathbb{1}}$ has more \enquote{types} with a low 
 frequency but less  with a high one, compared to the 
 distribution $n(f)$ of real sequence  types. This confirms that 
 the distribution of the TCR repertoire does not just result 
 from random selection but instead from enhancement and 
 suppression during V(D)J  recombination and thymus selection. 
Some \CDR sequences are chosen with a probability above the random average, thus they are enhanced. 
Others stay below the average, they are suppressed. We mark the crossing point in red, where \CDR sequences are neither suppressed nor enhanced. 
The empirical data (diamond shaped green marks) show a similar decline, compared to the generated \CDR sequences. 
Of course, this distinction remains only schematic, since 
the exact value of $N_{max}$ in the TCR repertoire is not known exactly and estimates deviate by several orders of magnitude. But it clearly shows that the histogram mainly reflects the selection effects from V(D)J recombination and Thymus selection.

\subsection{Network structure of TCR repertoires}
\label{network_structure}
The ample size and complexity of the TCR repertoire impedes the understanding of its structure and  biological implications.  The decades-old field of network
science provides a set of tools to overcome this hurdle. Hence, in the following section, we focus on basic 
network properties, namely the degree distribution, and the LCC of a TCR network. We employed our TCR network construction framework TCR-NET to create large networks, which form the base for this analysis. Nevertheless, also the 
calculation of network properties for a given network can require significant computational effort, 
and varies greatly between different network properties. Therefore, we will 
use different network sizes in dependence of the network properties that we want to determine, varying from several thousand \CDR sequences in computationally expensive properties such as the betweenness distribution or the assortativity to up to $8\times 10^5$ \CDR sequences for the robustness analysis. Doing that enables us to always use the largest networks possible for each analysis. 

\subsubsection{Assortativity and mean degree}
In Fig. \ref{fig:fig_2} (d) we analyzed the 
assortativity and the mean degree of the TCR network. 
We repeated the analysis of Fig. \ref{fig:fig_2}  (a) for $n=4$ TCR repertoires, but this time each of size $N = 8\times 10^{3}$ 
and selected all \CDR sequences (in total $N_{total}=32\times 10^3$) 
to construct the \enquote{total network} with $l_{max}=2$. 
Then we selected only the sequences, 
labeled as \enquote{public} and constructed the 
network with them, labelling it the \enquote{public network} 
as illustrated in Fig. \ref{fig:fig_2} (d), choosing $l_{max} = 2$ and having $N_{sim} = 364$ and $N_{exp} = 646$ empirical and generated  public sequences respectively. For generated sequences as well as experimentally obtained sequences, the public network has a similar or even lower assortativity but a higher 
mean degree than the total network, reflecting 
the denser structure. This indicates, that public 
sequences are more similar to each other, while 
private sequences cover a more diverse range of sequences for immune protection. 
\subsubsection{Betweenness centrality and neighbor degree distribution}
\label{neighbor}
Panel (e) of Fig. \ref{fig:fig_2} shows the normalized histogram of the betweenness centralities of nodes for a network constructed from $N=8\times 10^3$ sequences and with $l_{max} = 1$. 
There are a relatively small number of nodes with a 
high centrality compared to the majority of nodes with much lower centrality. 
In panel (f) of Fig. \ref{fig:fig_2} the average neighbor degree $\langle d_{nn}(d) \rangle$ 
(see section \ref{complex_networks}) is plotted 
in dependence of the degree $d$ of a node itself, with \NSeqPublicAnalysis 
and $l_{max} = 1$. 
We observe that a higher degree is associated with a higher average neighbor degree. 
The latter increases proportionally with the square root 
of the degree of the node itself (see  solid curves in panel (f) of Fig. \ref{fig:fig_2}) for generated, 
as well as empirical sequences. 

From the positive slope and the positive assortativity (for the latter see Fig. \ref{fig:fig_2} (d)) we conclude that the TCR network is an  assortative network. 
Thus, there are dense areas, where nodes are highly interconnected. On 
the other hand, the nodes with low degree are connected to neighbors 
which also have a low degree. 
This suggests that the immune system prepares against a certain group  of  epitopes by creating a dense 
net of interconnected TCRs while against more \enquote{unusual} antigens there is
only a relatively weak protection. 
  To visualize that, the inset 
 shows a TCR network of $N = 8\times 10^{3}$ \CDR sequences with $l_{max} = 1$, where the 
 radial position of each node is inversely proportional 
 to its degree, while the polar angles 
 are chosen randomly. Each node is colored depending on its average neighbor degree, yellow indicating a high value, while dark blue a low value. It can be observed, that the nodes in the center, i.e. having a high degree, are also connected to higher degree nodes, i.e. they have a higher average neighbor degree. 

A goal for the future would be to investigate, how the immune system 
responds to the aforementioned \enquote{unusual antigens}. Perhaps they are really so 
scarce that the necessity to prepare an immune  response does not arise.
It also remains unclear if TCRs in these empty areas 
are perhaps more versatile and able to 
respond to a more diverse range of antigens, differing among each other by a higher Levenshtein distance (higher degeneracy).
The observed assortativity is unusual, since many biological networks tend to be more disassortative~\cite{newman2002assortative}.
Nevertheless, positive degree assortativity 
is also found in other networks outside of the area of immunology, e.g. 
the \enquote{e-printarchive coauthorship network} from arxiv.org~ \cite{lee2006statistical}. They all form the aforementioned group of assortative networks.

\subsubsection{Robustness of TCR repertoires}
Spontaneous mutations in the TCR repertoire of humans and animals can alter their immune systems' 
response to antigens. 
Moreover, the average lifetime of T cells is estimated to be not more than 
a few weeks~\cite{westera2013closing}.  Therefore, the structure of the TCR repertoire changes constantly. Moreover,  diseases such as HIV can lead to a  reduction of T cells~\cite{philip2021cd8, hazenberg2000t}. This raises the question, how long characteristic structural properties stay preserved. 
We simulate a change in the repertoire with the removal of a TCR, thus the removal 
of a node, which can be caused, e.g., by apoptosis, lysis, or anergy 
of T cells.
To study the robustness of the TCR network, we then analyze how  certain TCR network properties change under node removal, which  is also commonly performed for other network 
types in graph theory~\cite{callaway2000network, schneider2013algorithm} and can be understood as a site percolation problem (see section \ref{transition}).

In Fig. \ref{fig:fig_2} (g) we show the normalized relative size (for definition see section \ref{complex_networks} and \ref{computing_LCC} under Methods) of the largest connected component (LCC) as a function of $1-p$, where $p := \frac{N_{r}}{N}$ denotes the fraction of removed nodes, $N_{r}$ the total number of removed nodes, and $N$ the total number of nodes in the network before any node removal. We chose the parameters  \NSeqRobustness and $l_{max} = 1$.
The relative size of the LCC continuously  declines during node removal. But even after removing half of all nodes ($p=0.5$), the relative size has decreased by less than twenty percent, indicating 
a strong robustness of the TCR network and low susceptibility to damages. The main cluster stays preserved. This  is in agreement with the results found 
in Ref.~\cite{miho2019large} for antibodies 
and necessary in order to maintain the 
immunological footprint of an individual. 
The empirical 
data show a similar robustness of the immune system, with only slight deviations. 
On the other hand, a loss of more than eighty percent of all nodes strongly damages the network. In contrast to other network types, such as Erd{\H o}s--R{\'e}nyi graphs~\cite{tishby2018revealing}, 
 the TCR network does not exhibit a phasetransition at some point $p_C<1$ where the LCC abruptly vanishes. 
Instead, a significant portion 
always remains, which continuously decreases toward zero when $1-p$ becomes close to 
zero, similar to the Barabási--Albert model~\cite{PhysRevLett.85.4626} (see section \ref{complex_networks} under Methods). 
This corresponds to the limes
\begin{equation}
    \lim\limits_{p \rightarrow 1} \textnormal{LCC} = 0
\end{equation}
The spontaneous jumps in the relative size of the LCC can be explained with a large connected component, only being connected to the rest of the LCC by one remaining edge. If this last edge is removed due to the removal of a node, then the size of the LCC can shrink by a large portion, since a large number of nodes is suddenly cut of.
In Fig. \ref{fig:fig_2} (h) we plot the normalized degree distributions (see section \ref{complex_networks} and \ref{normalization}) 
corresponding to the node-removal probabilities 
$p\in \{ 0.01, 0.49, 0.89, 0.95, 0.98\}$
for the same network parameters as in Fig. \ref{fig:fig_2} (g). A high $p$ is represented by a bright green line, a low $p$ by a dark green tone for generated sequences and analoguously in blue for empirical sequences.  The normalized frequency of each degree value is plotted in dependence of the respective degree as a histogram.
As expected, for empirical as well as generated sequences, the frequency decreases with higher $p$, i.e. higher number of removed nodes, but the 
negative slope also increases. The networks degree distribution shows similarity with an algebraic decline, which 
is also found in several  other  
graph types (see also section \ref{complex_networks}). 

\subsubsection{LCC transition}
A TCR recognises an antigen, e.g. from a virus or a cancer cell,   
 by detecting a characteristic site of the antigen, called {\itcmd epitope}, 
  which matches the {\itcmd paratope} of the TCR. 
Usually, the paratope is not specific to only one epitope of an antigen (for a detailed introduction to paratopes and epitopes see Ref.~\cite{e2003fundamental} chapter 4). During the early 2000s 
 it became concensus that there is a significant degeneracy 
 where each paratope can recognize multiple epitopes, then called a {\itcmd mimotopic array}~\cite{cohn2005degeneracy},  
 and each epitope can be recognized by multiple paratopes, referred to 
 as a {\itcmd paratopic clan}~\cite{cohn2005degeneracy}.
 Thus, the TCR repertoire does not need every paratope, i.e., every \CDR sequence, to be present because a paratope can recognize some neighbor epitopes too, as long as they remain sufficiently similar to the \enquote{ideal} epitope. We model the condition of sufficient similarity by requiring the Levenshtein distance between \CDR sequences to be smaller than or equal to a maximum distance $l_{max}$, as explained in the introduction. 
 We want to investigate 
 how small $l_{max}$ can be to guarantee full coverage, 
 which we define as given if the size of the LCC is of the same order of magnitude as the entire TCR repertoire ($LCC \sim N$).
 It should be noted that the Levenshtein distance is a very rough 
 measure of the similarity between two \CDR sequences, 
 which actually depends on molecular interactions. In some cases a small Levenshtein distance can lead to nonrecognition while in other cases an epitope can be recognized by a  sequence despite a high  Levenshtein distance to the ideal cognate sequence~\cite{wu2002two,aleksic2012different,reiser2003cdr3,wucherpfennig1995molecular,wucherpfennig2007polyspecificity}. Thus, the investigation using the Levenshtein distance
  should be always treated as an approximation.
 
  We will study the relative size of the LCC, as we did during the robustness analysis (see Fig. \ref{fig:fig_2} (g)), this time in dependence of the threshold value $l_{max}$, as shown in  Fig.  \ref{fig:fig_2} (i). 
The relative size  
 of the LCC 
is plotted in dependence of $l_{max}$ for different parameters 
$N\in  \{1\times 10^{3}, 4\times 10^{3},  8\times 10^{3}\}$.  
While at $l_{max} = 0$ the LCC covers a minority of nodes, 
a transition can be observed for increasing $l_{max}$,  
until the LCC covers the entire network and full coverage is given according to the previous definition. 
We can observe, that for higher $N$, this transition shifts left, i.e., toward lower thresholds $l_{max}$. 
We can interpret the position of this transition as an estimate of how big the disparity between a given \CDR sequence and an epitope's \enquote{ideal} cognate paratopes \CDR sequence can be in order
to still recognize the antigen and therefore enable an immune response. 
As larger the total number $N$ of TCRs, as 
smaller the requisite tolerance of each single node, thus as smaller the 
required degeneracy of each TCR for  
provoking a sufficient immune response. 
On the other hand, this indicates that smaller organisms may 
 need 
a wider responsiveness of each TCR to antigens due to their smaller TCR repertoire. 
This might be achieved through higher degeneracy.
Alternatively, it might also be possible that smaller 
organisms need to cover a less diverse sequence pool,
due to possible lower diversity of cell genes and 
possibilities of antigens to react with healthy cells. 

\subsection{Distinguishing antigens with network analysis}
\label{virus}
\begin{figure*}[htb]
    \centering
    \includegraphics[scale = 0.7]{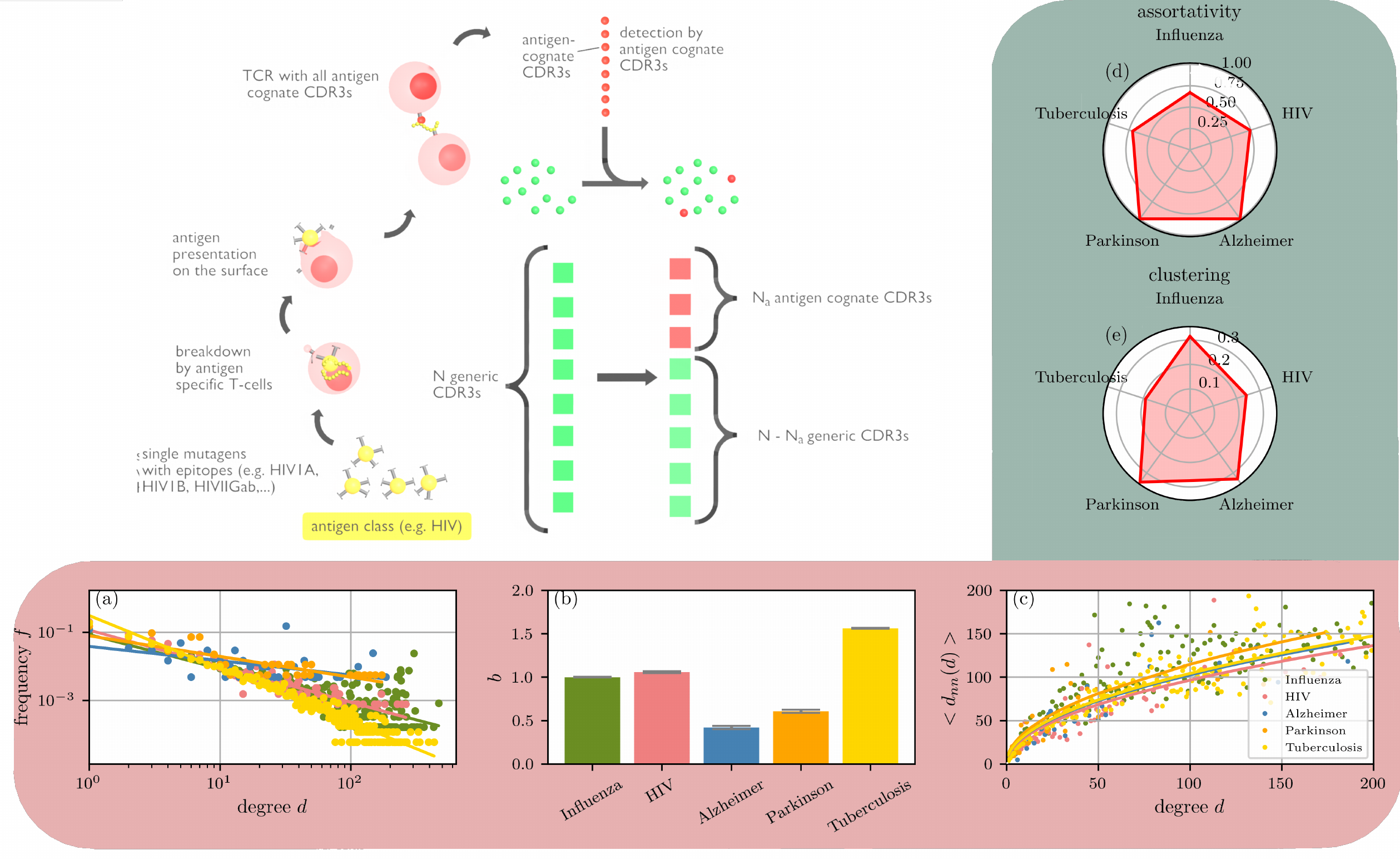}
    \caption{{\bf Network topology of different antigens}. The scheme in the upper left illustrates the insertion of antigen cognate \CDR sequences in the TCR network (explained in the text). 
Panel (a) shows the degree distribution of 
    different antigens ($N=4\times 10^5,  l_{max}=1$). Though they all resemble an algebraic decline, their decay exponents $b$ differ and are shown in panel (b).
Panel (c) shows the monotonically positive 
    average neighbor degree 
    distribution for the same network. 
Panel (d) and (f) show the radarplots of     
    assortativity and clustering respectively. 
    }
    \label{fig:fig_3}
\end{figure*}

 The paratope is the part of the TCR, that comes into contact with the antigen epitope (presented by an {\itcmd antigen presenting cell}) 
  and 
  mainly consists 
  of the three CDR regions within the variable domain of the TCR (for a details we refer to chapter 4 in Ref.~\cite{e2003fundamental}. 
  Since the diversity within a single individual predominantly arises from the CDR3 region, it suffices to represent a paratope by its CDR3 region, that can then be matched to the specific epitope, called the {\itcmd cognate} epitope. 
  This of course neglects cross reactivity  (for introduction in cross reactivity see Ref.~\cite{e2003fundamental})
   and hence should be treated as an approximation. 
  In a previous work a large database was created~\cite{tickotsky2017mcpas}, matching  
  \CDR chains to the corresponding antigen epitopes for different antigen families. 
  We use this data in order to represent antigens through their corresponding cognate \CDR sequence 
  and place them as a node in our TCR network.
  To do that, we begin with a normal list 
  of $N$ \CDR empirical sequences $\{\sigma_i|\ 0 < i \leq N\}=:\{\sigma_i\}$. Moreover, we define the list $\{\sigma_{a}|\ 0 < i \leq N_a, N_a \ll N\}=:\{\sigma_a\}$ as the list of $N_a$ \CDR sequences, that are cognate to at least one antigen from the antigen family $a$ (e.g. Influenza or Tuberculosis). 
  We choose the antigen families of the viruses Influenza, HIV, Tuberculosis
  as well as of Alzheimer and Parkinson, where, though not infectious diseases, the immune system plays a major role ~\cite{heppner2015immune, jevtic2017role,  kustrimovic2018parkinson}. 
  We now select all $N_a$ \CDR sequences from the 
  set $\{\sigma_{a}\}$ and replace the first $N_a$ sequences in $\{\sigma_i\}$ with them. We call the resulting set $\{\sigma_{i,a}\}$ and construct the network from it,  labeling it an {\it antigen specific network} or briefly {\it ASN}. 
  In this way we create one network for each antigen and 
  can compare the network properties of  antigen cognate nodes, inserted from $\{\sigma_a\}$ (abbreviated with {\itcmd ac nodes}). 
The network parameters are chosen to be $l_{max} = 1$, 
 $N=4\times 10^5$. Moreover, the values $N_A$ are 6051, 1291, 204, 177, 16663 for Influenza, HIV, Alzheimer, Parkinson, Tuberculosis respectively and therefore of a lower order of magnitude than $N$. 

Panel (a) of Fig. \ref{fig:fig_3} shows the normalized degree distribution 
of ac nodes in different ASNs. Though they share the common behavior of an algebraic decline, as in many other 
network types, 
their decline exponents differ. 
While Alzheimer slowest decline with a 
decay exponent of $b = -0.419\pm 0.018$, 
the degree distribution of Tuberculosis declines fastest with the exponent of $b = -1.563\pm 0.005$, as shown in Panel (b).  In contrast to the  
other investigated antigens, Influenza 
has an unusually large number of nodes with high degree,  
deviating from the algebraic curve. The given standard deviations should be taken with caution, since the fit relies on the assumption of an algebraic decline. If the degree distribution differs slightly from a perfect algebraic distribution, this would lead to additional deviations.
In panel (c) we show the average neighbor degree distribution, similar to panel (f) of Fig. \ref{fig:fig_2}, but only took ac  nodes into account. 
Again, we observe a square root curve, even with similar scaling throughout the different antigens, within 
the bounds of uncertainty. Therefore, also the subgraphs of ac nodes are assortative,  if we include edges between ac nodes and other nodes in the network.
Panel (d) and (e) show the assortativity and the clustering coefficient  
for different ASNs, only calculated from ac nodes. They deviate significantly throughout the antigens, making them a possible marker to distinguish antigens. 

It should be noted that a single antigen often carries 
multiple epitopes 
and thus can react to multiple TCRs. 
More databanks about cross reactivity would offer
 opportunities to perform further network analysis in the future.

\begin{figure}
\centering
\includegraphics[scale = 1.0]{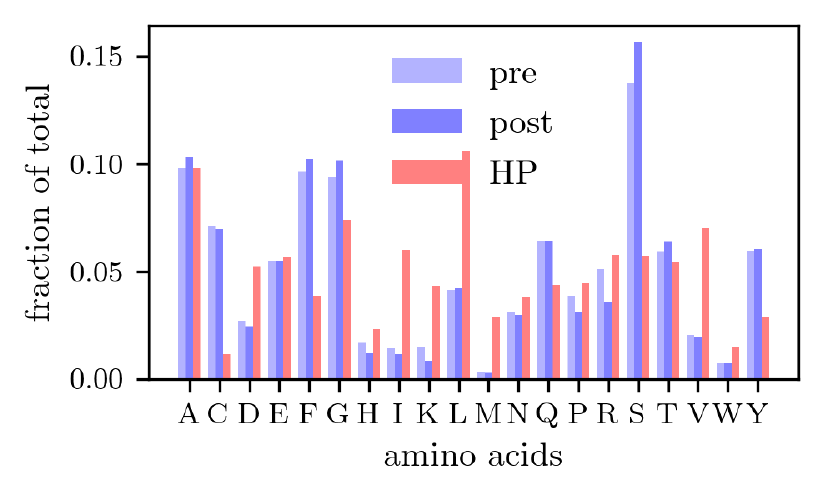}
\caption{{\bfseries Histogram of the a.a. frequencies.} 
We show the distribution of a.a. relative frequencies for a human TCR repertoire of size $N = 10^{6}$ before the thymus selection (\enquote{pre}) and 
after thymus selection (\enquote{post}), using SONIA generated \CDR sequences. While the distributions don't differ largely from each other, they both deviate significantly from the a.a. distribution in human plasma (\enquote{HP}). The amino acids are denoted by their standard one-letter abbreviations. 
}
\label{fig:aminoHist}
\end{figure}

\section*{Amino acid composition}
TCR receptors consist of amino acid sequences (a.a.s.). 
That leads to the question how the a.a.s. are  distributed. 
This has been analyzed for various
different tissues over many decades. 
Deviations in the a.a. composition can indicate
illnesses~\cite{mally1997changes}. 
Ref.~\cite{song2018recent} 
reviews the currently applied methods and different tissues that have been
analyzed, including human plasma. 
We compare the a.a. composition of the \CDR 
sequences with the composition of human plasma~\cite{song2018recent, mcmenamy1960studies} in Fig. \ref{fig:aminoHist}. 
It shows the normalized frequency of each a.a. within a sample of $N = 10^6$ 
 \CDR sequences from TCR$\beta$ chains, generated with SONIA. The distribution before thymus 
selection (light blue)
is 
similar to the distribution after the thymus 
selection (blue), with exception of K, P, R and S (we refer to amino acids, using their 1-letter symbols).
These sequences are significantly enhanced (S) or inhibited (K, P, R) 
by the thymus. 
Both distributions show much stronger differences, compared to 
the average a.a. distribution in human plasma (red), 
with exceptions of 
A, E, N, P, R and T, where the relative concentrations
  only differ by a relatively small amount. 
 It can be concluded that the \CDR sequences do not just resemble
 the average a.a. composition of human plasma but instead have a unique composition, which is 
 only changed slightly during thymus selection.

\section{Discussion}
 \label{conclusion}
 We created the package {\scshape TCR-NET}, which improves the performance compared to previous GPU based algorithms for Levenshtein distance calculation~\cite{balhaf2016using} by exploiting the specificities of \CDR sequences in the TCR repertoire. 
 Using the SONIA sequence simulation software package~\cite{sethna2020population}, we then  
 created a sequence pool in order to analyze the fraction of public and private 
 sequences. In coherence with previous analysis~\cite{elhanati2018predicting, quigley2010convergent, robins2009comprehensive, freeman2009profiling, benichou2012rep, six2013past, robins2010overlap, venturi2011mechanism, elhanati2014quantifying, zvyagin2014distinctive,  pogorelyy2017persisting, madi2014t}, it was found that the fraction of public sequences depends on the 
 definition of public, which still differs among different scientific publications. 
 Moreover, a higher repertoire size increases the share of public  sequences. 
  Experimental evidence suggests that the 
 final fraction of public sequences lies significantly above the theoretically predicted values resulting from pure stochasting modeling of V(D)J recombination~\cite{soto2020high}. 
 We could confirm a difference between generated \CDR sequences and empirical sequences from naive CD8 T cells throughout all investigated definitions of public. 
 Moreover, a comparison with a random distribution showed that the public sequences are not just the result of a stochastic overlap,  
 which could also occur in an entirely random distribution, but instead must derive from enhancement and suppression effects during V(D)J recombination and thymus selection.  
 
 In the next step, we investigated the network structure of the TCR repertoire.
 It was found that the average neighbor degree of a node increases proportionally to the square root of the degree of the node itself,  showing that the network is assortative. Nodes with many neighbors seem to be connected with each other in a 
 very dense center while some isolated nodes can be found far outside. 
 The LCC for TCR networks from SONIA-generated \CDR sequences as well as empirical sequences proved to be robust against large damages ($N_r \sim N$), indicating that the TCR repertoire can retain its charactaristic topology. 
 Of course the network properties depend significantly on  $l_{max}$. If $l_{max}$ increases for constant $N$, the size of the LCC increases as well until
  covering the entire network. The curve expresses a transition 
  behavior, where the transition shifts towards smaller values of $l_{max}$ for larger 
  $N$. 
  
  Moreover, we inserted \CDR sequences that correspond to certain antigens~\cite{tickotsky2017mcpas} into TCR networks and analyzed their network properties. 
  Especially Influenza exhibited an unusually high number of high degree nodes, deviating from the algebraic decline observed for the degree distributions of all investigated antigens. 
  Moreover, the clustering coefficients and assortativities differ between the  antigens.
  The neighbor degree distributions for all antigen specific networks increase monotonically, indicating an assortative network. 
  All in all this provides strong evidence, that antigen families can be distinguished by their network topology, especially the degree distribution,  clustering coefficient and assortativity. 
  
  Finally, we investigated the a.a. composition of generated human \CDR sequences  and found that it  deviates significantly from the average a.a. composition in human plasma. Moreover, there are also (though more slight) differences between the 
  compositions before and after thymus selection. This might help distinguishing samples and opens up perspectives to empirical  investigations on how 
  diseases or special conditions might change the a.a. composition in the TCR repertoire, as it 
  was done for other parts of the human body~\cite{engelen2001effects,mally1997changes}. 
  
  In conclusion, we studied the share of public \CDR sequences, the robustness of the immune system, the network structure, the relation to antigen  epitopes as well as the a.a. composition and related it to previous work. 
  Especially the application of network theory in TCR repertoire analysis is 
  a promising field, which still opens up huge spaces for further investigations and can help understanding the immune systems structure
  and responsiveness in more detail than it was previously possible. 
  Especially with the rise of personalized medicine~\cite{hamburg2010path} it will become necessary
  to understand the specific differences in the immune system between individuals. This will require a very detailed analysis of the complex topology of the TCR repertoire. Thus, especially with the rise of high-throughput sequencing and consequently the availability of more and 
  more data, the field of network theory is an increasingly helpful tool to 
  understand the TCR repertoire. 

\acknowledgements 
We would like to thank Prof. Dr. Markus Sigrist from ETH Zurich, Prof. Tom Chou from UCLA and Dr. B\"ottcher from University of Frankfurt for their support and helpful discussions.
 \section{Methods}
This section introduces some fundamental concepts and mathematical definitions used throughout the project, ranging from \CDR sequence generation to fundamentals on sets, statistics and network theory to details about our network construction framework, in particular with regard to the GPU implementation.   
\subsection{Generation of \CDR sequences} 
During V(D)J recombination, the probability that a certain TCR  sequence is formed by a recombination event $E$ can be approximated by the equation~\cite{sethna2020population,murugan2012statistical}
\begin{align}
\begin{split}
P_{\rm recomb}(E) & = P_V(V)P_{DJ}(D, J)\\
&\cdot P_{{\rm del}V}(d_V|V)P_{{\rm del}J}(d_J|J)P_{{\rm del}D}(d_D, d_D'|D)\\
&\cdot P_{{\rm ins}VD}(l_{VD})\prod\limits_{i=1}^{{\rm ins}VD}p_{VD}^{(2)}(x_i|x_{i-1})P({\rm ins}DJ)\\
&\cdot \prod\limits_{i=1}^{{\rm ins}DJ}p_{DJ}^{(2)}(y_i|y_{i-1})
\end{split}
\label{eq:Pscenario}
\end{align}
Here, $V$, $D$, and $J$ are gene sequences from the variable, diversity, and joining regions, respectively. $P_V(V)$ is the probability, that a certain sequence $V$ is chosen from the variable region. $P_{DJ}(D, J)$ is the joint probability, that the genes $D$ and $J$ are chosen from the diversity and joining region. $P_{{\rm del}V}\left(d_V| V\right)$ is the conditional probability that a number $d_V$ of bases is deleted at the end of the $V$ sequence. The probabilities $P_{{\rm del}D}\left( d_D, d_D'|D \right)$ and $P_{{\rm del}J}(d_J| J)$ are defined analogously, where $d_J$ is the number of bases that is deleted at the end of the $J$ section. Since the $D$ region is located at the center of a TCR sequence, $d_D$ and $d_D'$ correspond to the number of deleted bases at the two ends of the sequence. Moreover, $P_{{\rm ins}VD}(l_{VD})$ and $P_{{\rm ins}DJ}\left(l_{DJ}\right)$ are the probabilities that a certain sequence of length $l_{VD}$ or $l_{DJ}$ is inserted at the $V$-$D$ junction or $D$-$J$ junction, respectively. 
The conditional probability $p_{VD}^{(2)}(x_i | x_{i-1})$ is the probability that the base at position $i$ has the value $x_i$ given the value $x_{i-1}$ of the base at position $i-1$. This accounts for a nearest-neighbor dependency of the base selection. Analogously, $p_{DJ}^{(2)}(y_{i}|y_{i-1})$ is the probability for the base at position $i$ to have the value $y_i$, while the base at position $i-1$ has the value $y_{i-1}$. The total number of parameters is 2,856~\cite{murugan2012statistical}, and most of them are associated with the deletion probabilities.
\subsection{IGoR, OLGA, SONIA} 
\label{sonia}
During T-cell formation, two important processes have to be disentangled. One is the generation of a distribution of T-cell clones due to V(D)J recombination in the thymus. The second process is the selection of clones in the thymus based on their 
reactive behavior. Both processes can be described with mathematical models, embedded in software packages. The first process is captured by the IGoR package~\cite{marcou2018high}, which learns the model based on empirical data by adjusting the parameters based on the comparison of generated sequences and experimentally obtained reference sequences. The OLGA algorithm~\cite{10.1093/bioinformatics/btz035} increases the efficiency and thus performance of this task. SONIA is a software tool that additionally incorporates the second process, that is, thymus selection~\cite{sethna2020population}.
 \subsection{Histogram and relative frequency}
 \label{Histogram and relative frequency}
 May $\boldsymbol{x} = \left(x_1, x_2, ...,  x_n\right)$ be a vector of $n$ elements, each chosen from a set $\boldsymbol{\tilde{x}} = \{\tilde{x}_i\ |\ 0<i\leq \tilde{n}\}$ of $\tilde{n}$ different elements  $\tilde{x}_i$. 
 Then the  {\itcmd histogram function} $m(\tilde{x}\in \boldsymbol{\tilde{x}})$ (or short  {\itcmd histogram})  can be defined as
 \begin{equation}
     m(\Tilde{x}) = |\{i\in \mathbb{N}_{|\boldsymbol{x}|}:x_i = \Tilde{x}\}|, 
 \end{equation}
 where $|\cdot|$ denotes the {\itcmd cardinality} of a set. 
 We define the  {\itcmd relative frequency $f$} of an element $\Tilde{x}\in \boldsymbol{\tilde{x}}$ as
 \begin{equation}
     f(\tilde{x}) = \frac{ |\{i\in \mathbb{N}_{|x|}:x_i = \Tilde{x}\}|}{|\boldsymbol{x}|} = \frac{m(\tilde{x})}{|\boldsymbol{x}|}. 
 \end{equation}
 This normalization assures $\sum\limits_{\tilde{x} \in \boldsymbol{\tilde{x}}} f(\tilde{x}) = 1$. We also refer to the distribution of relative frequencies as the {\it normalized histogram}. 
 \subsection{Cartesian product}
 \label{Cartesian product}
 The {\itcmd Cartesian product} $C$ between two sets $A$ and $B$ is defined as 
 \begin{equation}
     C = A\times B = \{(a, b)|\ a\in A\ \textnormal{and}\ b\in B\}, 
 \end{equation}
 where $C$ is a new set, comprising tuples $(a, b)$ as elements. For instance,  
 the Cartesian product of the two sets $A = \{1, 2, 3\}$ and $B = \{\alpha, \beta, \gamma \}$ is 
 \begin{equation}
 \begin{split}
     C = \{(1, \alpha), (1, \beta),  (1&, \gamma), (2, \alpha), (2, \beta),  (2, \gamma), \\ 
     &(3, \alpha), (3, \beta), (3, \gamma)\}. 
  \end{split}
 \end{equation}
 
 The n-ary  {\itcmd Cartesian power} of a set $X$ is defined as 
 \begin{equation}
 \begin{split}
     X^{n} &= \underbrace{X\times X\times ... \times X}_{n} \\
     &= \{(x_1, x_2, ...,  x_n)\ |\  x_i\in X\ \forall i: 0<i\leq n\}. 
 \end{split}
 \end{equation}
 
 \subsection{Combination}
 \label{Combination}
 May $\boldsymbol{x}$ denote a set containing $N_{max}$ elements. 
 Moreover, $\boldsymbol{y}$ shall be a set of $N$ elements, each chosen from $\boldsymbol{x}$,  where the same element can be chosen multiple times, thus $N$ can be larger than $N_{max}$. Then the function $C_{\boldsymbol{p}}(N, N_{max}, \boldsymbol{x})$ shall be defined as the mapping from $\boldsymbol{x}$ to $\boldsymbol{y}$, where 
 $\boldsymbol{p} = (p_1, p_2, ..., p_{N_{max}})$  denotes the vector of probabilities $p_i$ to choose an element $x_i\in \boldsymbol{x}$. In the next section we will write  $C_{p}(N, N_{max})$ or $C_{p}$ for brevity and make the input set $\boldsymbol{x}$ implicit since we are more interested in the mapping function itself than in the concrete elements. 
 
 \subsection{Combination of amino acids}
 \label{Combination of amino acids}
 We represent each amino acid 
 with its corresponding 1-letter symbol. Then the 20 different amino acids (a.a.) form the set 
\begin{align*}
    \Sigma = \ &\{A, R, N, D, C, Q, E, G, H, I, L, \\
    &K, M, F, P, S, T, W, Y, V\}. 
\end{align*}
Thus, a sequence $\boldsymbol{\sigma}$ containing a number $l$ of a.a. is an element of the set  $\Sigma^{l}$ of all possible a.a. sequences with length $l$. The total set of all possible a.a. sequences with arbitrary length then can be defined as $\Sigma_{total} = \bigcup_{l=1}^{\infty}\Sigma^{l}$. In reality, not all of these a.a. sequences are possible during V(D)J recombination. Instead,  the choice reduces to a subset $\Sigma_{N_{max}} \subset \Sigma_{total}$ of $N_{max}$ possible a.a. sequences. Then an array of $N$ a.a. sequences, generated during V(D)J recombination, is an element of the set $\Sigma_{N_{max}, N} := \Sigma_{N_{max}}^{N}$ (the N-ary Cartesian power of $\Sigma_{N_{max}}$ - see section \ref{Cartesian product}). 

The combination $C_{\boldsymbol{p}}$ uses $N \in \mathbb{N}$ and $N_{max} \in \mathbb{N}$ as an input in order to generate an array of $N$ a.a. sequences $\{\boldsymbol{\sigma}\}_N:=\{\boldsymbol{\sigma}_i |\ 0<i\leq N\} \in \Sigma_{N_{max}}^{N}$. We view the vector $\boldsymbol{p}$ of weights of all possible a.a. sequences as a parameter instead of an input. Thus, the combination can be defined as  

\begin{align*}
C_{\boldsymbol p}: & \ \mathbb{N} \times \mathbb{N} & \rightarrow & \  \bigcup_{0<N<\infty}\ \bigcup_{0<N_{max}<\infty}\Sigma_{N_{max}}^{N} \\
   \ & \left(N, N_{max}\right) & \mapsto & \ \{\boldsymbol{\sigma}\}_N .
\end{align*}

\label{dynamic}

\subsection{Overview of network theory concepts} 
\label{complex_networks}

A graph (\ie, network) $G(V, E)$ is an ordered pair $(V,E)$, 
where $V$ is the set of nodes and $E\subseteq V\times V$ is the 
set of edges that connect pairs of 
nodes~\cite{newman2018networks}. In this work, we consider 
undirected and unweighted networks. The connectivity properties 
of such a network are described by the symmetric {\it adjacency matrix} $A$ with binary entries $0$ and $1$. If an edge connects two nodes $u,v\in V$, the corresponding adjacency matrix entry $A_{uv}=A_{vu}$ is 1 and 0 otherwise. For each node $u \in V$, the {\it degree} is given by the number of edges attached to that node, that is, $d(u)=\sum_{v\in V} A_{u 
v}=\sum_{v\in V} A_{v u}$. The {\it degree distribution} $p_d$ is the relative frequency of nodes with degree $d$ and defines the mean degree according to $\langle d \rangle=\sum_{d} d p_d$. The {\it clustersize distribution} is defined as the relative frequency of clusters with size $CC$. 
The {\it shortest path length} between a source node $s$ and a target node $t$ is the minimum number of edges that connect these nodes. If the shortest path length is finite for every node pair $(s,t)\in V\times V$, the network is {\it connected}. A network is {\it fully connected} (or complete) if every node is connected to every other node. A {\it subgraph} of $G(V,E)$ is a network $G_{sub}(V',E')$ over a subset $V'\subseteq V$ of nodes and the subset $E'\subseteq E$ of all edges that connect nodes in $V'$. The {\it largest connected component} (LCC) is the largest connected subgraph. 
The {\it relative size} of a subgraph $G_{sub}$ of $G$ 
is defined as the quotient $\frac{|G_{sub}|}{|G|}$, where $|G_{sub}|$ and $|G|$ denote the number 
of nodes in the subgraph and graph respectively (also called the {\it size} of a graph). 

The {\it degree assortativity} $r$ is defined as~\cite{newman2002assortative}
\begin{equation}
r = \frac{\sum_{kl}kl\left(e_{kl} - q_k q_l\right)}{\sigma_q^2},
\end{equation}
where $q_k=(k+1)p_{k+1}/\langle k \rangle$ is the distribution of {\it excess degree}. The {\it average neighbor degree} of a network is 
\begin{equation}
\langle d_{nn} \rangle = \sum\limits_{d'}d'P\left(d'| d\right),
\end{equation}
where $d$ is the degree of a node, $d'$ is the degree of a neighboring node, and $P(d'|d)$ is the conditional degree distribution. 
The {\it betweenness centrality} of a graph is defined as 
\begin{equation}
g(v) = \sum\limits_{s\neq v\neq t} \frac{\sigma_{st}(v)}{\sigma_{st}},
\end{equation}
where $\sigma_{st}$ denotes the number of shortest paths from a node $s$ to a node $t$ and $\sigma_{st}(v)$ is the number of shortest paths, which cross the node $v$.

\subsection{Percolation transition}
\label{transition}
An interesting phenomenon in many networks, e.g. Erd{\H o}s--R{\'e}nyi graphs, is the percolation phase transition.   
We distinguish {\itcmd bond percolation} and {\itcmd site percolation}. While the former process describes the change of network properties during removal of edges, the latter describes changes under node removal. 
Thus, we randomly choose and remove a fraction $p$ of all edges or nodes respectively, while observing how a network property, e.g. the LCC, changes for higher $p$. In the following we focus on bond percolation. 
Erd{\H o}s--R{\'e}nyi graphs, as many other graph types, exhibit a transition at a $p_c < 1$ where the relative size of the LCC vanishes~\cite{tishby2018revealing}. This value lies at $p_{c,ER} = \frac{1}{\langle k \rangle}$, where $\langle k\rangle$ refers to the average node degree in the Erd{\H o}s--R{\'e}nyi graph. In contrast, Barabási--Albert graphs do not exhibit this transition for a  $p_c$ below one~\cite{PhysRevLett.85.4626}. Instead, the LCC continuously shrinks and only vanishes when the number of edges goes to 0. Thus, $p_{c,BA} = 1$.  

\subsection{Normalization}
\label{normalization}
Throughout the analysis we employed different normalization procedures, which are defined below. 
To normalize the sharing number distribution, shown in the panels (a) and (b) of Fig. \ref{fig:fig_2}, the frequency $f(k)$ of \CDR sequences with a certain sharing number $k$  will 
be divided by the total number $N$ of sequences: 
\begin{equation}
    f_{norm} = \frac{f(k)}{N}
\end{equation}
For the normalization procedure of the betweenness distribution and the degree distribution, we refer to the histogram normalization, described in section \ref{Histogram and relative frequency}. 
\subsection{Computing LCC}
\label{computing_LCC}
To normalize the LCC in the robustness analysis in panel (g) of Fig. \ref{fig:fig_2}, we first divide the number of nodes in the LCC (denoted by LCC$(p)$) by the number of 
 nodes in the network after node removal, given by $(1-p)N$, i.e., 
\begin{equation}
    \textnormal{LCC}_{norm,1}(p) = \frac{\textnormal{LCC}(p)}{(1-p)N}
\end{equation} 
In a second step we divide by the relative size of the LCC at $p=0$, i.e., when 
no nodes were removed:  
\begin{equation}
\textnormal{LCC}_{norm, 2}(p) = \frac{\textnormal{LCC}_{norm,1}(p)}{\textnormal{LCC}_{norm,1}(0)}. 
\end{equation}
In panel (i) of Fig. \ref{fig:fig_2} we don't remove any nodes, only perform the first normalization and therefore plot $\textnormal{LCC}_{norm,1}(p=0) = \frac{\textnormal{LCC}(0)}{N}$. 

\subsection{Averaging}
\label{averaging}
In order to reduce fluctuations, we can  average over multiple samples to calculate a distribution $y = f(x)$. 
May each sample, referred to via index $i\in \{1, 2, ..., M\}$, where $M$ denotes the number of samples, have its unique distribution $f_i(x)$. Then the average over all distributions is calculated as
\begin{equation}
    \bar{f}(x) = \frac{1}{M}\sum\limits_{i=1}^{M}f_i(x).
\end{equation}

\subsection{Levenshtein distance}
\label{levenshtein_distance}

The {\itcmd Levenshtein distance} can be used to compare two strings $a$ and $b$, each having arbitrary length, and is defined as~\cite{levenshtein1966binary}

\begin{align}
\resizebox{.9\hsize}{!}{
$\textnormal{lev}_{a, b}\left( i, j\right) = \left \{ 
\begin{array}{cc}
\max \left( i-1, j-1 \right)  &\textnormal{if $i=1$}\\ &\textnormal{or $j=1$}\\
\min \left\{ 
\begin{array}{c}
\textnormal{lev}_{a,b} \left( i-1, j \right) + 1\\
\textnormal{lev}_{a,b} \left(i, j-1\right) + 1\\
\textnormal{lev}_{a,b} \left(i-1, j-1\right) + 1_{\left(a_i \neq b_j\right)}
\end{array} 
\right.
& \textnormal{else}
 \end{array}
\right . $
}
\label{eq:levenshtein}
\end{align}

with 

\begin{equation}
1_{\left(a_i \neq b_j\right)} = \left\{
\begin{array}{cc}
0&\textnormal{if $a_i = b_j$}\\
1&\textnormal{ otherwise.}
\end{array}
\right.
\end{equation}

$a_i$ denotes the $i$th entry of the string $a$. Likewise, $b_j$  corresponds to the $j$th entry of $b$. 
If $a$ and $b$ are equal in length, the Levenshtein distance is also called {\itcmd Hamming distance} and corresponds to the number of elements in which both strings differ. E.g. if $a=\text{ACCA}$ and $b=\text{ACGG}$, both strings only differ at positions three and four. Hence, their Hamming distance corresponds to two. 
Since \eqref{eq:levenshtein} is recursive, we performed the computation of the Levenshtein distance iteratively, based 
on a straight-forward algorithm. 
\subsection{TCR network}
\begin{figure*}
\includegraphics[scale=0.9]{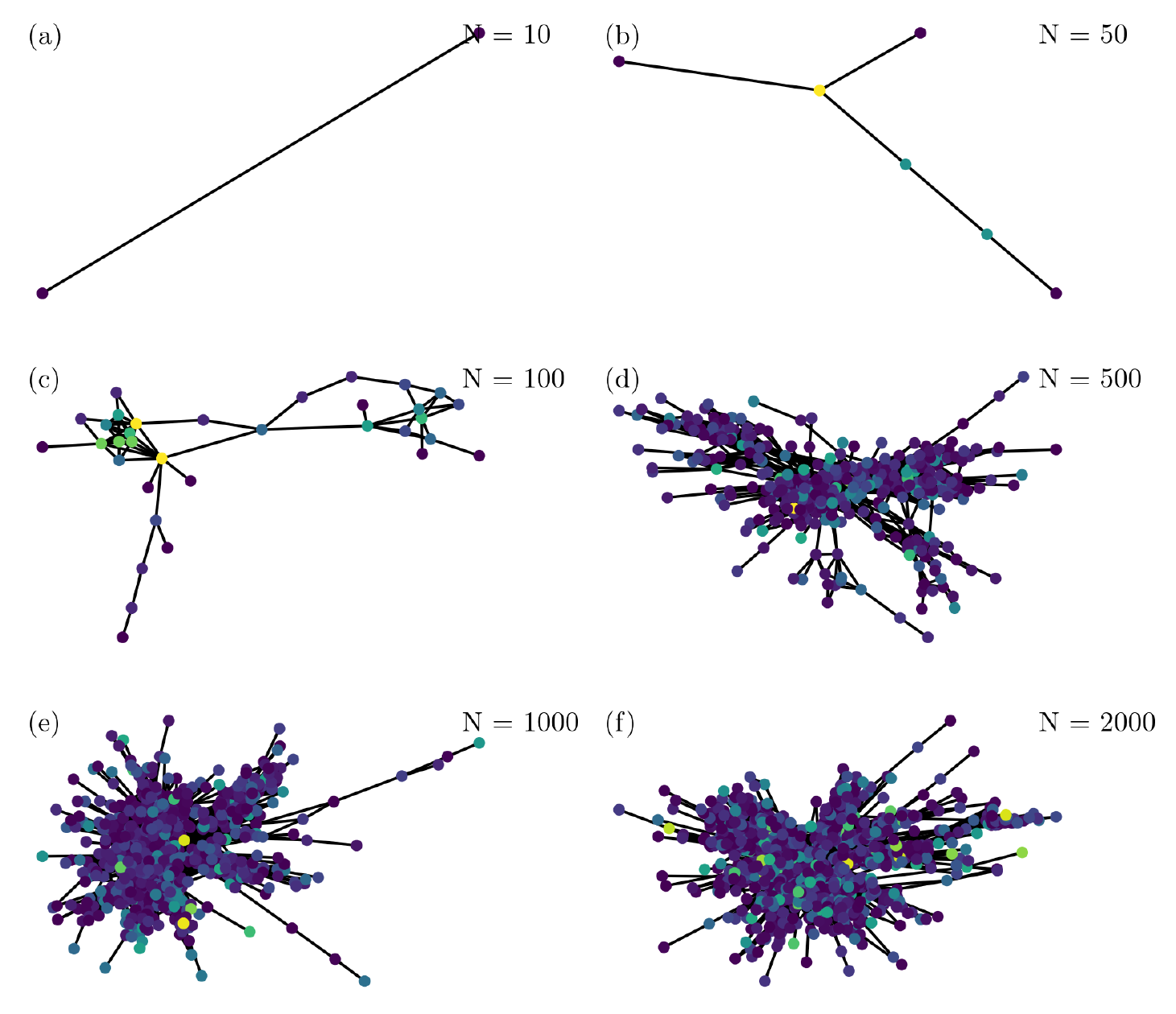}
\caption{ 
In panels (a) to (f) 
we draw the largest connected component (LCC) of TCR networks with different sizes $N$ but constant $l_{max} = 6$. In each graph, the color indicates the degree, where the node with maximum degree occurs in yellow and the nodes with lowest degree  in dark blue. 
}
\label{fig:graph}
\end{figure*}
We define a TCR network as a graph $G$ where each \CDR sequence $\boldsymbol{\sigma}$ is represented by a node, also called TCR node (see graphs of different sizes in Fig. \ref{fig:graph}). If and only if the Levenshtein distance $l$ between two \CDR sequences does not exceed a threshold value $l_{max}$ (thus $l \leq l_{max}$), the two corresponding nodes are connected via an edge. 

 \subsection{Adjacency matrix}
 Let's consider a graph $G$ with $n$ nodes, connected via a set $E$ of edges. Each edge $e\in E$ can be represented via a tuple of two numbers $(n_1, n_2) \in \mathbb{N}_n\times \mathbb{N}_n$, where $n_1$ and $n_2$ denote the indices of the two connected nodes. Since the edges have no direction ($G$ is an {\itcmd  undirected} graph), switching the two entries of a tuple yields the same edge, i.e. $(i, j) \cong (j, i)\ \forall {i, j \in \mathbb{N}_n}$. 
 The edges can be represented by a symmetric matrix, referred to as the {\itcmd  adjacency matrix}, where

 \begin{align*}
     (i, j) \mapsto \begin{cases} 0, \ &\textnormal{if} \ (i, j)\notin E \ \textnormal{and} \ (j, i)\notin E \\
     1, \ &\textnormal{if} \ (i, j)\in E \ \textnormal{or} \ (j, i)\in E.\end{cases}
\end{align*}
 
 For example, let us consider a  graph $G$ with five nodes and the edges
 \begin{equation*}
     E =  \{(1, 2), (1, 3), (2, 5), (4, 5)\}.
 \end{equation*}
 Then, the adjacency matrix becomes
 \begin{equation}
     \begin{pmatrix}
     0&1&1&0&0\\
     1&0&0&0&1\\
     1&0&0&0&0\\
     0&0&0&0&1\\
     0&1&0&1&0\\
     \end{pmatrix}_{.}
 \end{equation}
 For larger graphs, this matrix increases in scale and the number of matrix entries (and thus the computational effort to construct the matrix) increases quadratically with the $n$. 

\subsection{GPU-based network generation}
\label{sec:GPU_algorithm}
\begin{figure*}
    \centering
    \includegraphics[scale=1.0]{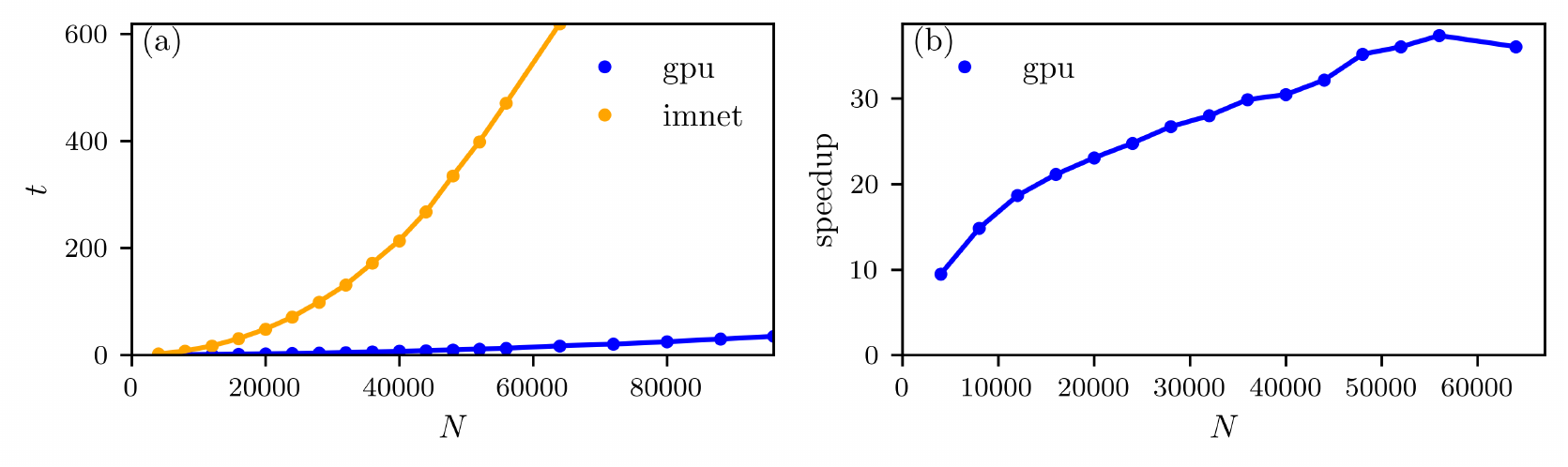}
    \caption{{\bf Performance of TCR-NET.} (a) The runtime to calculate Levenshtein distances between all possible pairs of $N$ \CDR sequences using the CPU-based algorithm {\sc imNet}~\cite{miho2019large} and our GPU-based implementation TCR-NET with $l_{\rm max} = 1$. (b) The computational speedups associated with using TCR-NET for Levenshtein distance calculations. All calculations were performed on one core of an Intel i7-7700 CPU and an Nvidia GTX 1660 GPU. The GPU speedup indicates a lower bound of the number of CPU cores necessary to outperform GPU calculations for the same number of sequences.}
    \label{fig:performance}
\end{figure*}
The advantage of GPUs is their ability to parallelize processes, what  
can be utilized to calculate the Levenshtein distance between sequences and with that the entries of the adjacency matrix, more efficiently, as done previously~\cite{balhaf2016using}. 
We used a Nvidia GTX 1660, with  $t=1408$  kernels and 6 GB GDDR6.  For large networks its memory does not suffice to store the entire adjacency matrix.  
In order to overcome this restriction, we employ the strategy of {\itcmd tiling}, 
 where we divide the total adjacency matrix into blocks, 
which are calculated consecutively. To further reduce the memory consumption, each block is saved as a sparse matrix. 
The GPU speedup is compromised by the 
lower operating frequency of GPU kernels compared to CPU kernels and the 
transfer time between the GPU and the CPU RAM. 
 All this has to be accounted for while implementing an efficient GPU 
 algorithm. 
 
 Knowing that most \CDR sequences are only 8 to 11 a.a. long  
 and the relative frequency of sequences declines sharply 
 with higher length, we reserved memory space for only 30 a.a. per sequence on the GPU
  in order to limit the computational workload. 
  Moreover, there are only 20 different common a.a. Each a.a. is saved as a number. When they are converted, these numbers are not saved as {\sc int} (32 bit) but instead as {\sc byte} (8 bit), 
  which occupies only a quarter of the memory on the GPU.
  We use numba to perform distance calculations on the GPU and  determine the sparse adjacency matrix.
  All these processes have 
  the goal of improving the computing efficiency.
In order to further reduce the computation time the iterative calculation of the levenshtein distance between two a.a. sequences, will abort, as soon as the smallest intermediate result exceeds the maximum allowed levenshtein distance $l_{max}$. The algorithm to calculate the levenshtein distance is shown as pseudocode in Listing \ref{listing1}. 
 
 In Fig. \ref{fig:performance} we compare the 
 performance of {\sc TCR-NET} with the previously published high-performance framework {\sc  imnet}~\cite{miho2019large}. TCR-NET beats {\sc imnet} on a normal computer with a single core of an Intel i7-7700 CPU by approximately one to two orders of magnitude, with a higher speedup for large networks. Of course {\sc imnet}'s performance has an advantage on large CPU clusters, where the high number of kernels  enables an efficient use of the pyspark parallelization. Moreover, TCR-NET only 
 calculates efficiently if the number of edges constitutes merely a small fraction of all possible edges, which is usually true for a TCR network, if $l_{max}$ is low.   
 
 We employed {\sc Python} as the main programming language and used  numba and pycuda for GPU optimization. For complex network computation, the package {\sc networkx} was  employed in order to supplement the packages {\sc numpy} for mathematical computation and {\sc matplotlib} for 
plotting. 
The {\sc imnet} package employs igraph and R  
\cite{rcore2016r}
\cite{csardi2006igraph}. 
With the {\sc SONIA} package 
 a.a. strains could be generated directly from the commmand line~\cite{sethna2020population}.
 
The full source code and additional information are publicly available on GitHub~\cite{GitHub}.
 \lstset{language=Python}

 \begin{lstlisting}[caption = {Pseudocode for the calculation of the Levenshtein distance. The source code and additional information is publicly available on GitHub~\cite{GitHub}.}, basicstyle=\small, frame = single]
t = len(string_1)
s = len(string_2)

dnew = zeros(s);
dprev = zeros(s);

# info: calculating 
#     the Levenshtein distance
for (i = 1; i < len(t); i++):
    dprev[0] = i-1;
    dnew[0] = i;
    
    for (j = 1; j < len(s); j++):
        if(string1[i-1] == string2[j-1]):
            substitutionCost = 0
        else:
            substitutionCost = 1
        
        dnew[j] = 
            min(dnew[i-1])+1, dprev[i]+1, 
                dprev[i-1] + 
                substitutionCost)
        
        if (dnew[j] < min_val):
            min_val = dnew[j]
        
    for (j = 1; j < len(s); j++):
        dprev[j] = dnew[j];
        
    if (min_val > max_ld):
        break;
# info: checking, if the Levenshtein  
#     distance is in the desired 
#     range; c_value represents the
#     entry in the adjacency 
#     matrix 
if dnew[len(s)-1] < max_ld:
    c_value = 1
else:
    c_value = 0
    
 \end{lstlisting}
\label{listing1}
\bibliographystyle{apsrev4-1}
\bibliography{net}
\end{document}